\documentclass[12pt,preprint]{aastex}

\usepackage{booktabs}
\usepackage{flushend} 

\newcommand{\FeII}{[\ion{Fe}{2}]}
\newcommand{\SII}{[\ion{S}{2}]}

\newcommand{\kms}{km~s$^{-1}$}

\newcommand{\degree}{^{\circ}}
\newcommand{\Ha}{H${\alpha}$}

\newcommand{\myemail}{hyoh@kasi.re.kr}

\slugcomment{}

\shorttitle{IGRINS Spectroscopy of jets around LkH$\alpha$ 234}
\shortauthors{Oh et al.}

\begin{document}


\title{IGRINS Near-IR High-Resolution Spectroscopy \\
of Multiple Jets around LkH$\alpha$ 234\footnote{This paper includes data taken at The McDonald Observatory of The University of Texas at Austin.}}


\author{Heeyoung Oh\altaffilmark{1,2}, Tae$-$Soo Pyo\altaffilmark{3,4}, In$-$Soo Yuk\altaffilmark{1}, Byeong$-$Gon Park\altaffilmark{1,2}, Chan Park\altaffilmark{1}, Moo$-$Young Chun\altaffilmark{1}, Soojong Pak\altaffilmark{5}, Kang$-$Min Kim\altaffilmark{1}, Jae Sok Oh\altaffilmark{1}, Ueejeong Jeong\altaffilmark{1,7}, Young Sam Yu\altaffilmark{1},
Jae$-$Joon Lee\altaffilmark{1}, Hwihyun Kim\altaffilmark{1,6}, Narae Hwang\altaffilmark{1}, Kyle Kaplan\altaffilmark{6}, \\Michael Pavel\altaffilmark{6}, Gregory Mace\altaffilmark{6}, Hye$-$In Lee\altaffilmark{5}, Huynh Anh Nguyen Le\altaffilmark{5}, Sungho Lee\altaffilmark{1} and Daniel T. Jaffe\altaffilmark{6}}
 
%
%


\altaffiltext{1}{Korea Astronomy and Space Science Institute, 776 Daedeok-daero, Yuseong-gu, Daejeon 305-348, Korea. {\myemail}}
\altaffiltext{2}{Korea University of Science and Technology, 217 Gajeong-ro, Yuseong-gu, Daejeon 305-350, Korea.}
\altaffiltext{3}{Subaru Telescope, National Astronomical Observatory of Japan, 650 North A'ohoku Place, Hilo, HI 96720.}
\altaffiltext{4}{School of Mathematical and Physical Science, SOKENDAI (The Graduate University for Advanced Studies), Hayama, Kanagawa 240-0193, Japan.}
\altaffiltext{5}{School of Space Research and Institute of Natural Sciences, Kyung Hee University, 1732 Deogyeong-daero, Giheung-gu, Yongin-si, Gyeonggi-do 17104, Korea.}
\altaffiltext{6}{Department of Astronomy, University of Texas at Austin, Austin, TX 78712.}
\altaffiltext{7}{Center for Atmospheric Remote Sensing (CARE), Kyungpook National University, 80 Daehakro, Bukgu, Daegu, 702-701, Korea.}


\begin{abstract}
We present the results of high-resolution near-IR spectroscopy toward the multiple outflows around the Herbig Be star Lk{\Ha} 234 using the Immersion Grating Infrared Spectrograph (IGRINS). Previous studies indicate that the region around Lk{\Ha} 234 is complex, with several embedded YSOs and the outflows associated with them. In simultaneous H$-$ and K$-$band spectra from HH 167, we detected 5 {\FeII} and 14 H$_{2}$ emission lines.
We revealed a new {\FeII} jet driven by radio continuum source VLA 3B. Position-velocity diagrams of H$_{2}$ 1$-$0 S(1) $\lambda$2.122 $\micron$ line show multiple velocity peaks. The kinematics may be explained by a geometrical bow shock model. We detected a component of H$_{2}$ emission at the systemic velocity (V$_{LSR}$ $=$ $-$10.2 {\kms}) along the whole slit in all slit positions, which may arise from the ambient photodissociation region. Low-velocity gas dominates the molecular hydrogen emission from knots A and B in HH 167, which is close to the systemic velocity, {\FeII} emission lines are detected at farther from the systemic velocity, at V$_{LSR}$ $=$ $-$100 $-$ $-$130 {\kms}. We infer that the H$_{2}$ emission arises from shocked gas entrained by a high-velocity outflow. Population diagrams of H$_{2}$ lines imply that the gas is thermalized at a temperature of 2,500 $-$ 3,000 K and the emission results from shock excitation. \end{abstract}



\section{Introduction} \label{sec:intro}
Lk{\Ha} 234 (V373 Cep) is a B7-type Herbig Be star in the NGC 7129 cluster \citep[D $=$ 1.25 kpc,][]{Shevchenko1989}. The star is $\sim$ 8.5 M$_{\sun}$, $\sim$ 1700 L$_{\sun}$ \citep{Hillenbrand1992}, and has an age of $\sim$ 10$^{5}$ years \citep{Strom1972,Fuente2001}. Lk{\Ha} 234 is bright in far-IR bands and is surrounded by reflection nebulae \citep{Bechis1978,Bertout1987}. It is one of the objects in the catalog of embedded clusters \citep{Lada2003}, and is a well-known water maser emission source \citep{Cesarsky1978,Rodriguez1987,Trinidad2004,Marvel2005,Bae2011}.

Lk{\Ha} 234 region is complex, with multiple outflows from different YSOs.
\citet{Edwards1983} discovered a large-scale outflow in J = 1$-$0 $^{12}$CO, which is mostly redshifted with low-velocity ($\sim$ 10 \kms), to the northeast of Lk{\Ha} 234. Shocked H$_{2}$ flows are also detected in the CO outflow region \citep{Eisloffel2000}. Mid-IR spectroscopy showed that this shocked H$_{2}$ is collisionally excited by a J-type shock \citep{Morris2004}. 

\citet{Ray1990} found a blueshifted optical {\SII} jet (HH 167), which extends more than 30$\arcsec$ with a P.A. of 252$\degree$. \citet{Ray1990} argue that this jet is the counterpart of the red CO lobe. \citet{Schultz1995} showed that there is near-IR H$_{2}$ emission at the position of the {\SII} jet, and \citet{Cabrit1997} confirmed an H$_{2}$ jet in high angular resolution images. The H$_{2}$ jet coincides with knots A, B and C of the {\SII} jet \citep{Ray1990}. \citet{Kato2011} noticed two jet-like features in their stellar coronagraphic images in the J$-$ and H$-$bands. One of the jet-like features is coincident with a ``green blob'' in the middle of the reflection nebula shown in their JHK color composite image. The position of the green blob is in agreement with the {\FeII} emission in \citet{Schultz1995}.
In addition, \citet{McGroarty2004} suggested a parsec-scale jet over $\sim$ 22$\arcmin$ ($\sim$ 8 pc) as projected on the sky, including HH 815 $-$ 822 and HH 103 A.
In Figure \ref{fig:sourcemap}, we show the source candidates and the axes of outflows on the images taken from \citet{Kato2011} and \citet{Cabrit1997}. Also, Table 3 in \citet{Kato2011} established the association of sources and source labels from previous studies, which helps the understanding of this region.

Polarimetric observations at 2 $\micron$ by \citet{Weintraub1994} revealed the existence of an embedded young stellar companion 3$\arcsec$ northwest of Lk{\Ha} 234. Their polarization map also showed that the companion is the most likely illuminating source of the reflection nebula. \citet{Cabrit1997} detected a 10 $\mu$m source (IRS 6) at the position coinciding with the center of the polarization vectors in \citet{Weintraub1994}. \citet{Polomski2002} also detected a mid-IR source NW at the position of IRS 6. Since the axis of the H$_{2}$ jet (P.A. = 227$\degree$) is not aimed at IRS 6, the possible existence of another embedded source was suggested. IRS 6 is not driving the H$_{2}$ jet, but it was regarded as the source of the CO/{\SII} outflow \citep{Cabrit1997,Fuente2001}. Millimeter observations by \citet{Fuente2001} revealed a new embedded source, FIRS1-MM1 $\sim$ 4$\arcsec$ northeast of Lk{\Ha} 234. FIRS1-MM1 lies within 1$\arcsec$ the jet axis, so it is suggested as the source of the H$_{2}$ jet \citep{Fuente2001}.

Many radio observations have shown strong continuum emission near Lk{\Ha} 234 \citep[e.g.,][]{Wilking1986,Skinner1993,Tofani1995}. 
The strongest radio source, VLA 3, is associated with IRS 6. \citet{Trinidad2004} showed that the VLA 3 is a binary system (3A and 3B) having thermal radio jets.
In the 1.3 cm and 3.6 cm continua and H$_{2}$O maser observations, the water maser sources around another radio source VLA 2 are tracing the axis of {\SII} outflow. The proper motion of H$_{2}$O masers showed bipolar motions centered on VLA 2 along the axis of {\SII} outflow. These suggested that VLA 2 is the driving source of {\SII} outflow \citep{Trinidad2004,Torrelles2014}. \citet{Kato2011} detected NW 1 and 2 at the positions of IRS 6 and VLA 2, respectively in a mid-IR observation.

In Figure \ref{fig:sourcemap}, the positional uncertainties are less than 0\farcs2 and 0\farcs05 for the radio sources and the mid-IR sources (NW 1 and 2). For IRS 6, the cross indicates an uncertainty. We note that an additional error could be caused by astrometric matching between different observations, which is less than 0\farcs5.

From the previous studies \citep{Cabrit1997,Fuente2001}, we can say that the {\SII} outflow is driven by VLA 2 (=NW 2) and that FIRS1-MM 1 is well aligned with the near-IR H$_{2}$ jet axis. Currently there is no evidence of Lk{\Ha} 234 itself as an outflow driving source.

A large fraction of stars form in binary or multiple star systems \citep{Duquennoy1991,Connelley2008a,Connelley2008b,Reipurth2014,Pineda2015}. For all stars, outflows and mass-accretion are an essential part of their formation and early evolution \citep{Hartigan1995}. While studies of single star-disk systems have brought much knowledge about outflows and mass-accretion from young stellar objects, the complicated nature of multiple-star systems makes it more difficult to understand their characteristics in detail. High-resolution spatial-kinematic analysis toward outflows in young multiple systems can help us to understand their complicated velocity distribution and physical conditions.
In this paper, we report the first high-resolution near-IR spectroscopic observations of the multiple outflows around Lk{\Ha} 234 using the recently developed Immersion GRating INfrared Spectrograph \citep[IGRINS,][]{Yuk2010,Park2014}. We detected emission lines from {\FeII} and in ro-vibrational transitions of H$_{2}$ including {\FeII} $\lambda$1.644 $\micron$ and H$_{2}$ 1$-$0 S(1) $\lambda$2.122 $\micron$ lines. From our result, we suggest a newly confirmed {\FeII} outflow, possibly driven by VLA 3B. In Section \ref{sec:observation}, we describe the observations and data reduction. Section \ref{sec:result} shows the obtained position-velocity diagrams (PVDs) and line ratios. We discuss the characteristics of the {\FeII} and H$_{2}$ emission, and the driving sources of multiple outflows in Section \ref{sec:discussion}. We summarize the results in Section \ref{sec:summary}.

\section{Observation and Data Reduction} \label{sec:observation}

The data were obtained during a commissioning run of IGRINS mounted on the 2.7m Harlan J. Smith Telescope at the McDonald Observatory of the University of Texas. The date of observation was July 13, 2014 (UT). IGRINS covers the whole H$-$ (1.49 $-$ 1.80 $\mu$m) and the K$-$bands (1.96 $-$ 2.46 $\mu$m) simultaneously, with a high spectral resolution R ($\lambda/\Delta\lambda$) of $\sim$ 40,000 ($\Delta$$v$ $=$ 7.5 {\kms} with $\sim$ 3.5 pixels). Each of H$-$ and K$-$band cameras uses a (2k $\times$ 2k) HAWAII-2RG array for a detector \citep{Jeong2014,Oh2014}. The pixel scale along the slit, which becomes larger in higher orders was 0\farcs24 $-$ 0\farcs29 pixel$^{-1}$. The slit size was 1\farcs0 (W) $\times$ 14\farcs8 (L). Auto-guiding was performed during each exposure with a K$-$band slit-viewing camera equipped with a (2k $\times$ 2k) HAWAII-2RG array (pixel scale = 0\farcs12 pixel$^{-1}$). The guiding uncertainty on average was smaller than 0\farcs4. The seeing condition was about 1\farcs1 at K$-$band.

The observations were made at three slit positions covering LkH$\alpha$ 234 and HH 167. In Figure \ref{fig:sourcemap}, the slit positions are marked on the JHK color composite and H$_{2}$ narrow-band images. The white rectangles represent slit positions 1, 2, and 3 (hereafter SP 1, 2, and 3). The position angle (P.A.) of the slit was 256$\degree$ in all positions. We confirmed the slit positions and P.A. by matching the positions of stars on the slit-view images and the 2MASS Ks image. The total on-source integration time was 600 s for each slit position. The sky frames were obtained with the same exposure time as on-source frames. We observed HR 6386 (K $=$ 6.52, SpT $=$ A0V) as a standard star. Th-Ar and halogen lamp frames are acquired for wavelength calibration and flat-fielding, respectively.

We performed the basic data reduction with the IGRINS Pipeline Package\footnote{The IGRINS Pipeline Package is downloadable at \url{https://github.com/igrins/plp}. (doi:10.5281/zenodo.18579).} (PLP). The PLP performs sky subtraction, flat-fielding, bad pixel correction, aperture extraction, and wavelength calibration. To make position-velocity diagrams, we conducted additional processes on the data using IRAF (the Image Reduction and Analysis Facility). We resampled the spectra with a uniform wavelength interval per pixel along the dispersion direction by the TRANSFORM task (linearization). After we removed the Brackett series absorption lines from the standard star spectra using the SPLOT task, we did the wavelength sensitivity correction and flux calibration with standard star spectra. The telluric line correction was performed with the TELLURIC task. We used the BACKGROUND task to subtract the continuum emission from Lk{\Ha} 234.

\section{Result} \label{sec:result}

In Figure \ref{fig:sourcemap}c, the Y = 0$\arcsec$ position along the slit is indicated with a dash-dotted line, which crosses to the position of the Lk{\Ha} 234 in SP 1 and perpendicular to the P.A. of the slits. The notation of knots A, B, and C are from \citet{Ray1990}. Although the inclination angle of the jet is not known, the positional change on the sky plane of the knots is supposed to be small ($<$ 1\arcsec) due to large (1.25 kpc) distance from us.
Figure \ref{fig:profiles} shows line profiles of all detected 5 {\FeII} emission lines and 8 selected H$_{2}$ emission lines toward knot A at SP 1, 2, and 3.
In Figure \ref{fig3}, we show the position velocity diagrams of the H$_{2}$ 1$-$0 S(1) $\lambda$2.122 $\micron$ and {\FeII} $\lambda$1.644 $\micron$ lines at the three slit positions. The line profiles at 0\farcs5 interval along the Y direction are shown in Figure \ref{fig4}.

\subsection{Line profiles of {\FeII} and H$_{2}$  emission} \label{sec:profiles}
In Figure  \ref{fig:profiles}, the fluxes are integrated over knot A ($-$6\farcs5 $<$ Y $<$ $-$0\farcs5).
The {\FeII} lines at SP 1 and 2 show similar profiles with a peak velocity ($v_{peak}$) of $\sim$ $-$110 {\kms} and a shoulder toward lower velocity. At SP 3, only {\FeII} $\lambda$1.644 $\micron$ line is detected.
At each slit position, all H$_{2}$ lines show similar profiles .
At SP 1 and 3, the H$_{2}$ emission lines have asymmetric line profiles. At SP 1, the profiles peak at V$_{LSR}$ $\sim$ $-$11 {\kms} and show blueshifted wings. The profiles at SP 3 are asymmetric relative to the peak velocity (V$_{LSR}$ $\sim$ $-$13 {\kms}), showing a gradual slope on the red side of the line. At SP 2, they show double peaks at V$_{LSR}$ = $-$30 and $-$5 {\kms}.

The {\FeII} and H$_{2}$ emission profiles differ significantly; H$_{2}$ emission lines are strong at low-velocity, close to the systemic velocity of V$_{LSR}$ = $-$10.2 {\kms} \citep{Liu2011}, while {\FeII} emission lines are prominent at higher velocity. In the following section, we compare {\FeII} $\lambda$1.644 $\micron$ and H$_{2}$ 1$-$0 S(1) $\lambda$2.122 $\micron$ emission lines at each slit position in detail.

\subsection{PVDs of H$_{2}$ 1$-$0 S(1) $\lambda$2.122 $\micron$ and {\FeII} $\lambda$1.644 $\micron$ emission} \label{sec:2122and1644}

\subsubsection{Slit Position 1 (SP 1)} \label{SP1}

In Figure \ref{fig3}, contours and color gradients represent H$_{2}$ and {\FeII} emission, respectively.
SP 1 includes the southern part of knot A and the center of knot B, as shown in Figure \ref{fig:sourcemap}. The H$_{2}$ emission in knot A shows multiple intensity peaks. We marked two strong peaks as A1 and A2. They are at Y = $-$3\farcs5 and $-$4\farcs8, and their peak velocities are $\sim$ $-$25 and $-$11 {\kms}, respectively. A weak, but high$-$velocity ($-$110 \kms) emission is also detected at Y = $-$5\farcs0, at a similar position as knot A2. Velocity widths at half intensity ($v_{FWHM}$) are $\sim$ 37 and 23 {\kms} for A1 and A2, respectively. The $v_{peak}$ of A2 is very similar to the systemic velocity (i.e., $-$10.2 \kms). A plateau extended from knot A is shown at Y = $-$5\farcs3 and V$_{LSR}$ = $-$50 {\kms}. 

We detected a bright, high-velocity {\FeII} knot at Y = $-$4\farcs7 with a $v_{peak}$ $\sim$ $-$113 {\kms}. The position and velocity of this {\FeII} knot are almost consistent with high-velocity H$_{2}$ emission at Y = $-$5\farcs0 mentioned above. The {\FeII} emission also shows an extended velocity component which peaks at $-$73 {\kms}. The FWZI of the {\FeII} profile which contains both high-velocity and extended components is $\sim$ 130 {\kms}. From their double-peak line profile (Figure \ref{fig4}), we estimated the velocity widths and peak intensities by a multiple-gaussian fitting. $v_{FWHM}$ are 33 and 41 {\kms} for high-velocity and extended components, respectively. The peak intensity level of the extended component is about 3 times weaker than that of the bright peak ($\sim$ 40$\sigma$).

At Y $\sim$ $-$6\farcs8, both H$_{2}$ and {\FeII} emission show peaks at $-$130 \kms, which is the highest velocity in PVD. The position of this fast peak is located on the boundary region between knot A and knot B. We detected no counter component of low velocity in the H$_{2}$ emission.

Knot B reveals three different velocity peaks in H$_{2}$ emission. They show an increase in $|$Y$|$ with larger $ |v_{peak}|$ values. Their positions are Y = $-$9\farcs2, $-$9\farcs4, and $-$10\farcs2, and $v_{peak}$ are $-$13, $-$57, and $-$118 {\kms}, respectively. All three peaks have the same, narrow velocity widths of $\sim$ 23 $\pm$ 1 {\kms}. The lowest velocity emission at $-$13 {\kms}, which is close to the systemic velocity, is 8 times and 6 times brighter than those of the $-$57 and $-$118 {\kms} peaks, respectively.

In {\FeII}, two weak peaks are detected at Y = $-$9\farcs4 and $-$ 9\farcs9 with a $v_{peak}$ of $-$59 and $-$120 {\kms}. The $v_{FWHM}$ is wider (47 {\kms}) in its higher velocity peak than the slower peak (24 {\kms}). The positions and velocities of these peaks are consistent with $-$57 and $-$118 {\kms} peaks in H$_{2}$ emission. No {\FeII} emission corresponds to the position of the $-$13 {\kms} peak in H$_{2}$.

The {\FeII} emission usually arises from a similar shock region as the optical {\SII} emission \citep{Hamann1994}. The spectroscopy of {\SII} $\lambda\lambda$6716, 6731 {\AA} in \citet{Ray1990} showed $V_{LSR}$ of $-$79 and $-$89 {\kms} for knots A and B (velocities are converted from their heliocentric velocity). The peak velocities obtained from our {\FeII} observations are different from {\SII}. Knot A shows $\sim$ $-$113 {\kms} and knot B is detected in multiple velocities of $-$59 and $-$120 {\kms}. Although the intensity of knot B in {\SII} is comparable to that of knot A, {\FeII} in knot B is about 6 times weaker than knot A. One similarity between {\SII} and {\FeII} spectra is that both show no emission at the systemic velocity.

The differences between H$_{2}$ 1$-$0 S(1) $\lambda$2.122 $\micron$ and {\FeII} $\lambda$1.644 $\micron$ emission and their discrete multiple peaks will be discussed in detail in Section \ref{sec:discussion}.

\subsubsection{Slit Position 2 (SP 2)} \label{SP2}

SP 2 lies $\sim$ 1\farcs2 northwest of SP 1 (Figure \ref{fig:sourcemap}). This position includes the central region of knot A in H$_{2}$ emission. The H$_{2}$ and {\FeII} emission in this position are strongest among the three slit positions. SP 2 grazes knot B at Y $<$ $-$6\farcs8. 

In H$_{2}$ emission, knot A shows multiple velocity peaks at Y = $-$3\farcs5 and $-$5\farcs2 with $v_{peak}$ $\sim$ $-$30 and $-$5 {\kms} (Figure \ref{fig3}). Since their positions are different from A1 and A2 in SP 1, we marked them as A1$\arcmin$ and A2$\arcmin$. The $v_{peak}$ of A1$\arcmin$ is slightly faster than A1,  and A2$\arcmin$ is redshifted compared to the systemic velocity. The velocity widths of A1$\arcmin$ and A2$\arcmin$ ($\sim$ 34 and $\sim$ 25 {\kms}) are similar to those of A1 and A2, within an error range of $\pm$ 3 {\kms}. Weak emission at a high-velocity of $-$110 {\kms} is also detected at $-$4$\farcs$7. It is noticeable that the redshifted H$_{2}$ emission (V$_{LSR}$ $>$ $-$10.2 {\kms}) from knot A is extended to $\sim$ $+$30 \kms.

The {\FeII} emission from knot A is similar to that of SP 1 in velocity, showing double-velocity peaks. It shows a high-velocity peak of $-$112 {\kms} at Y= $-$4\farcs3 and an extended line profile toward lower-velocity. The peak of the extended component shows a $v_{peak}$ of $-$78 {\kms} and is $\sim$ 2.5 times weaker than the high-velocity peak. Their velocity widths estimated by a multiple-gaussian fitting are $\sim$ 32 and 46 {\kms}, respectively. The FWZI of this {\FeII} emission is the same with that of SP 1 ($\sim$ 130 {\kms}). {\FeII} peak in SP 2 is $+$0\farcs4 in Y from the corresponding peak in SP 1. This may be caused by the different P.A. of the {\FeII} outflow and our slits.

Weak H$_{2}$ emission ($\sigma$ $\sim$ 5) is detected at Y = $-$6$\farcs$8 and V$_{LSR}$ = $-$130 {\kms}, which coincides in position and velocity with the highest velocity peak of H$_{2}$ and {\FeII} at SP 1.

At Y $<$ $-$6\farcs8, a part of knot B shows a similar shape to that of SP 1 in H$_{2}$ emission: strong around the systemic velocity, high velocity peaks and a blueshifted wing. However, there is no relevant peak feature around the systemic velocity. {\FeII} emission is not detected in knot B.

\subsubsection{Slit Position 3 (SP 3)} \label{SP3}
At this position, we obtain a spectrum of the northwest part of knot A, while the slit does knot cover any of the emission from knot B (Figure \ref{fig:sourcemap}).
Knot A shows two peaks, as in the previous two slit positions. Both peaks lie near the systemic velocity and we mark them as A1$\arcsec$ and A2$\arcsec$. A1$\arcsec$ peaks at Y = $-$3\farcs35 and shows a less negative peak velocity ($-$12 \kms) than A1 and A1$\arcmin$, with a velocity width of 36 {\kms}. A2$\arcsec$ shows multiple velocities with peaks at $-$5 and $-$13 {\kms} at Y $\sim$ $-$5\farcs0. The velocity widths of the two peaks are 23 {\kms} and $\sim$ 41\kms, respectively.
In SP 3, an extension of a redshifted component at knot A in the H$_{2}$ emission is more noticeable than in SP 2. It is most extended at Y = $-$4\farcs5 and the velocity reaches over $+$40 \kms.

{\FeII} emission in SP 3 is much weaker than in SP 1 and 2, by a factor of $\sim$ 10 and $\sim$ 25, respectively.

\subsection{{\FeII} Line Ratios} \label{sec:feii}

Figure \ref{fig:feiiPVD} shows the PVDs of all detected {\FeII} emission from SP 1, SP 2 and SP 3. It includes the $\lambda$1.533, $\lambda$1.600, $\lambda$1.644, $\lambda$1.664, and $\lambda$1.677 $\micron$ lines. All five lines are detected at SP 1 and 2, while SP 3 only shows emission in $\lambda$1.644 line. In SP 1 and 2, the morphologies and velocities are almost the same in all of the lines in each slit position. One exception is in SP 1, where the highest velocity peak at V$_{LSR}$ $\sim$ $-$130 \kms, Y = $-$6\farcs8 and weak emission at Y = $-$9\farcs4 and $-$9\farcs9 are only detected in $\lambda$1.644 $\micron$. Measured line fluxes normalized to the {\FeII} $\lambda$1.644 $\micron$ line are listed in Table \ref{tbl:feiiflux}. Since the positions and velocities of peaks in all {\FeII} lines are consistent, the fluxes are estimated within the regions at Y $\sim$ 4\farcs7 $\pm$ 0\farcs4, V$_{LSR}$ $\sim$ $-$113 $\pm$ 10 {\kms} and at Y $\sim$ 4\farcs3 $\pm$ 0\farcs4, V$_{LSR}$ $\sim$ $-$112 $\pm$ 10 {\kms} in SP 1 and SP 2, respectively. The line fluxes in SP 1 and SP 2 are very similar.

Near-IR {\FeII} emission lines can be used as effective tracers of electron density ($n_{e}$) in stellar outflows \citep{Nisini2002,Pesenti2003,Takami2006}. \citet{Nisini2002} developed a non-LTE model applicable to jets from YSOs, considering the first 16 fine structure levels of the Fe$^{+}$ ion. The flux ratios between lines obtained in our observations are relatively independent of gas temperature, because the lines are all from the fine structure level of $a^{4}D$ term, having excitation energies in the range of 10,000 $-$ 12,000 K \citep{Nisini2002}. To investigate the electron density, we show the line ratios superposed on the model grids from \citet{Nisini2002} in Figure \ref{fig:Ne}. In SP 1 and 2, the electron densities (log $n_{e}$) averaged from the three ratios of $\lambda$1.600, $\lambda$1.533, $\lambda$1.677 $\micron$ lines against the $\lambda$1.644 $\micron$ line are 4.0 $^{+~0.7}_{-~0.6}$ and 4.1 $^{+~0.9}_{-~0.6}$, respectively. The uncertainty in $\lambda$1.644 $\micron$ / $\lambda$1.533 $\micron$ ratio is larger (Figure \ref{fig:Ne}), due to residuals after a subtraction of the OH sky emission in $\lambda$1.533 $\micron$ line. The $n_{e}$ of knot A we obtained from the optical {\SII} doublet ratio \citep{Osterbrock1989} is about 0.3 $\times$ 10$^{4}$ cm$^{-3}$, which is smaller than that estimated from near-IR ion emission lines here. This result is consistent with the tendency in \citet{Nisini2002}. It indicates that the {\FeII} lines are able to probe a denser region in the jet because they have a higher critical density ($\sim$ 10$^{4}$ cm$^{-3}$) than that of the {\SII} ($\sim$ 10$^{3}$ cm$^{-3}$).

The electron densities of 0.5 $-$ 3.2 $\times$ 10$^{4}$ cm$^{-3}$, obtained here are similar to or smaller than those estimated from {\FeII} emission from T-Tauri stars and intermediate-mass stars in \citet{Hamann1994}. In addition, these are also slightly lower than values from HH outflows from Class 0$-$I, which range from 10$^{4}$ to 10$^{5}$ cm$^{-3}$ \citep{Nisini2002,Takami2006}.

\subsection{H$_{2}$ Line Ratios} \label{sec:H2}

Table \ref{tbl:H2flux} lists the line fluxes of all 14 emission lines detected in these observations.
Fluxes are estimated in each peak position in Figure \ref{fig3}, and normalized to the H$_{2}$ 1$-$0 S(1) $\lambda$2.122 $\micron$ line.
Figure \ref{fig:H2PVD} shows the PVDs of 8 selected H$_{2}$ emission lines. In all slit positions, the different H$_{2}$ emission lines show similar velocity features, as in {\FeII}.
In all H$_{2}$ lines at SP 1,  2, and 3, we detected emission from both knot A and knot B which are bright at low-velocity. The three high-velocity components at Y = $-$6\farcs8, $-$9\farcs4, and $-$10\farcs2 with $v_{peak}$ of $-$130, $-$13, and $-$118 {\kms} shown in $\lambda$ 2.122 $\micron$ are also detected in $\lambda$2.034, $\lambda$2.223, $\lambda$2.407, and $\lambda$2.424 $\micron$ lines. In $\lambda$2.407 $\micron$ at SP 2, we detected a weak (5$\sigma$ level) emission which is consistent with a peak at Y = $-$4\farcs7 with $-$110 {\kms} in $\lambda$2.122 $\micron$. 

Figure \ref{fig:H2pop} shows H$_{2}$ population diagrams calculated from the fluxes in Table \ref{tbl:H2flux}.
The level populations of hydrogen molecular gas estimated from ro-vibrational transitions allow us to study the excitation state and temperature of the gas \citep{Black1976,Gautier1976,Beckwith1978,Hasegawa1987}. We calculated the upper level column density from observed line flux after a reddening correction. The extinction can be estimated using the ratios of the emission lines originated from the same upper levels. In our calculation, we used the ratios of three pairs of 1$-$0 S(1) / 1$-$0 Q(3), 1$-$0 S(0) / 1$-$0 Q(2), and 1$-$0 S(2) / 1$-$0 Q(4). We adopted the transition probabilities from \citet{Turner1977}. We apply the extinction law, A$_{\lambda}$ = A$_{\rm V}$(0.55$\micron$/$\lambda$)$^{1.6}$ \citep{Rieke1985}. In our estimation, A$_{\rm H}$ and A$_{\rm K}$ are very small ($<$ 0.1). We used A$_{\rm V}$ = 3.4 which is an intermediate from the range of 3.1 $-$ 5.1 used in previous studies \citep{Hillenbrand1992,Hernandez2004,Liu2011}.

For the data in Figure \ref{fig:H2pop}, we applied weighted least-square fitting to estimate rotational temperature (T$_{rot}$) from an upper level of $v$ = 1 (circle) and 2 (square), considering 1/$\sigma^{2}$ for weights, where $\sigma$ is the uncertainty in the level population. We excluded one transition from $v$ = 3 (triangle) from the fit due to its large uncertainty. Solid and dashed lines represent the fitting lines for $v$ = 1 and 2, respectively.

In SP 1, T$_{rot}$ of all peaks estimated in $v$ = 1 is 2500 $\pm$ 200 K. The temperatures in $v$ = 2 show uncertainty, because we have only three data points for the fitting. T$_{rot}$ of peaks A1 and B in $v$ = 2 are similar, 3200 $\pm$ 1500 K. In A2, T$_{rot}$ in $v$ = 2 is 2500 $\pm$ 1100 K which is close to the value from $v$ =1.
In SP 2, T$_{rot}$ in A1$\arcmin$ is almost the same as in A1, in both $v$ = 1 and 2. Peak A2$\arcmin$ shows a higher T$_{rot}$ than A2 in $v$ = 1 and 2, but the differences are within the uncertainties. In SP 3, T$_{rot}$ from A1$\arcsec$ and A2$\arcsec$ in $v$ = 1 are higher than those of SP 1 and 2; with T$_{rot}$ $>$ 2700 $\pm$ 250 K. In $v$ = 2, T$_{rot}$ is 3000 $\pm$ 1500 K for both A1$\arcsec$ and A2$\arcsec$. From SP 1 to 3, T$_{rot}$ in $v$ = 1 shows a small increase, but within uncertainties. For $v$ = 2, there is no clear trend with slit position change. In all three slit positions, in both $v$ = 1 and 2 the derived T$_{rot}$ is $\sim$ 2500 $-$ 3000 K. This means that the gases are thermalized in a single ro-vibrational temperature, because both level populations follow same T$_{rot}$. We note that the temperature estimation differs when we consider different A$_{\rm V}$ in the flux correction. It becomes $\sim$ 200 K lower and less than 100 K higher for the cases of A$_{\rm V}$ $=$ 0 and 5.1, respectively, compared to the estimation using A$_{\rm V}$ = 3.4. We also note that the populations of ortho and para transitions are aligned in a single line. This indicates that the emission lines are not from the fluorescent H$_{2}$ by UV excitation \citep[e.g.,][]{Hasegawa1987}.

Thermalization has been observed in the shocked gas in the outflows from low- and high-mass stars (e.g, HH 111, HH 240 $\&$ 241 in \citet{Nisini2002}, Orion KL peak1 in \citet{Beckwith1983}).
We can test the properties of shock with a population diagram by comparison with model data. \citet{Rosenthal2000} have investigated various shock models to prove a shock nature in Orion Peak 1, and they showed that the combining of shock models could explain their observational results except at the highest excitation level. Since we assume the presence of a bow shock in this region (see Section \ref{sec:discussion}), we plotted the bow shock model with $v_{shock}$ = 100 {\kms} from \citet{Smith1991} in Figure \ref{fig:H2pop} (dotted lines). This model is more consistent with fitting lines from $v$ = 2 than those from $v$ = 1. On the other hand, it is hard to discriminate the type of shock from our result because there is no significant difference between various shock models in the energy levels of 0.5 $-$ 2 $\times$10$^{4}$ K \citep{Rosenthal2000}.

In Table \ref{tbl:H2ratio}, we listed the H$_{2}$ line ratios at all peak positions that are sensitive to the type of shock. The ratios in the models of C-, J-type shocks and UV pumping are taken from \citet{Beck2008}. The ratios in our observation are similar to those of outflows from T-Tauri stars \citep{Beck2008}. The 2$-$1 S(1) / 1$-$0 S(1) ratio is close to C-type shock in all cases, indicating similar shock conditions along outflow. The other three ratios do not help discriminate between different shocks, due to the large uncertainties. There are no significant differences at all three slit positions. We note that the 1$-$0 S(1) / 1$-$0 Q(1) ratios in our data are about 1.5 $-$ 2.5 times higher than ratios from all models. 1$-$0 Q(1) $\lambda$2.407 $\micron$ emission is affected by a deep telluric absorption at $\sim$ $-$40{\kms} (`$\earth$' symbols in Figure \ref{fig:H2PVD}), so it may not be reliable.

\section{Discussion} \label{sec:discussion}

\subsection{{\FeII} and H$_{2}$ Emission}\label{sec:4.1}

As shown in Section \ref{sec:result}, the H$_{2}$ 1$-$0 S(1) $\lambda$2.122 $\micron$ and {\FeII} $\lambda$1.644 $\micron$ emission lines have are very different velocity structure. The H$_{2}$ emission is dominant at low-velocity while {\FeII} emission dominates at high-velocity. This trend was found in outflows from both CTTS and Class 0$-$I YSOs \citep [e.g.,][]{Pyo2002,Davis2003,Takami2006}. The trend makes sense in the context of an interpretation where the shocked H$_{2}$ emission arises from gas entrained by a high-velocity outflow \citep{Pyo2002}.

In our PVDs, we detected H$_{2}$ emission at the systemic velocity along the whole slit, in all three slit positions. This emission probably originates from ambient molecular hydrogen gas in this region, which corresponds to the presence of far-UV radiation by producing a PDR \citep{Morris2004}.
However, taking account of the H$_{2}$ lines ratios (Table \ref{tbl:H2ratio}), we should not rule out the excitation by shock as the emission source. We note that, in this case, the velocity width at the top of the PVDs should be larger than current value ($\sim$10 \kms), because the angle of the outflow axis may be smaller than 45$\degree$ with respect to the line of sight according to our interpretation in bow shock features described below.

Discrete multiple velocity components revealed from the H$_{2}$ emission in Figure \ref{fig3} are the typical high- and low-velocity features that arise from the bow shock in outflows \citep{Hartigan1987,Davis2003,O'Connell2004}. If the inclination angle between the bow shock axis and the line of sight is in a range of $\sim$ 0$\degree$ $-$ 45$\degree$, the emission from the apex and the wing of the bow are seen as separate high- and low-velocity peaks.
The weak H$_{2}$ and strong {\FeII} emission at high-velocity (V$_{LSR}$ $\sim$ $-$120 {\kms}) show good agreement in position and velocity, indicating that the {\FeII} emission lines usually arise from bow apex, where the jet velocity is faster than surrounding region \citep{Hartigan1987,Davis2003}.
The velocity difference of $\sim$ 100 {\kms} between the high- and low-velocity peaks in the H$_{2}$ lines is similar to that observed in the jet of L1551-IRS 5 \citep{Davis2003}, which has an inclination angle of $\sim$ 45$\degree$ with respect to the line of sight. In knot B at SP 1, we see three discrete velocity peaks in H$_{2}$ where the velocity of the peak increases with increasing $|$Y$|$. The more blueshifted peaks lie farther  from the star. The total offset along the outflow is $\sim$1\farcs15, after consideration of the different angle between our slit (256$\degree$) and the outflow axis (227$\degree$). One possible interpretation for this position shift is a spatial difference between the faster gas in the bow apex and the slower portion in the bow wing. We cannot rule out, however, the possibility that this shift is caused by distinct knots originally located in different positions.

In contrast to the blueshifted gas seen at all slit positions, the redshifted component (V$_{LSR}$ $>$ $-$10.2 {\kms}) seen in H$_{2}$ at SP 2 and 3 seems to arise from another, spatially unresolved outflow. Since the slit positions are pointing at the blue lobe of the jet (see Section \ref{sec:intro} and \ref{sec:observation}), at locations where the blueshifted material is well-separated in phase space from any ambient emission, the detection of redshifted emission implies the existence of a distinct outflow, separated from the high-velocity blueshifted jet. This outflow may have an axis almost parallel to the plane of the sky, with a wide outflow opening angle, because it shows both blue and redshifted components at low-velocity ($-$50 {\kms} $<$ V$_{LSR}$ $<$ $+$50 {\kms}).
For a further discussion, we need additional clues on the position and orientation of the redshifted component, which would need to come from further observations with high spectral resolution and with higher spatial resolution than we can obtain with our present instrument/telescope combination.

\subsection{Driving Sources of Multiple Outflow}

\begin{figure}
\epsscale{0.8}
\plotone{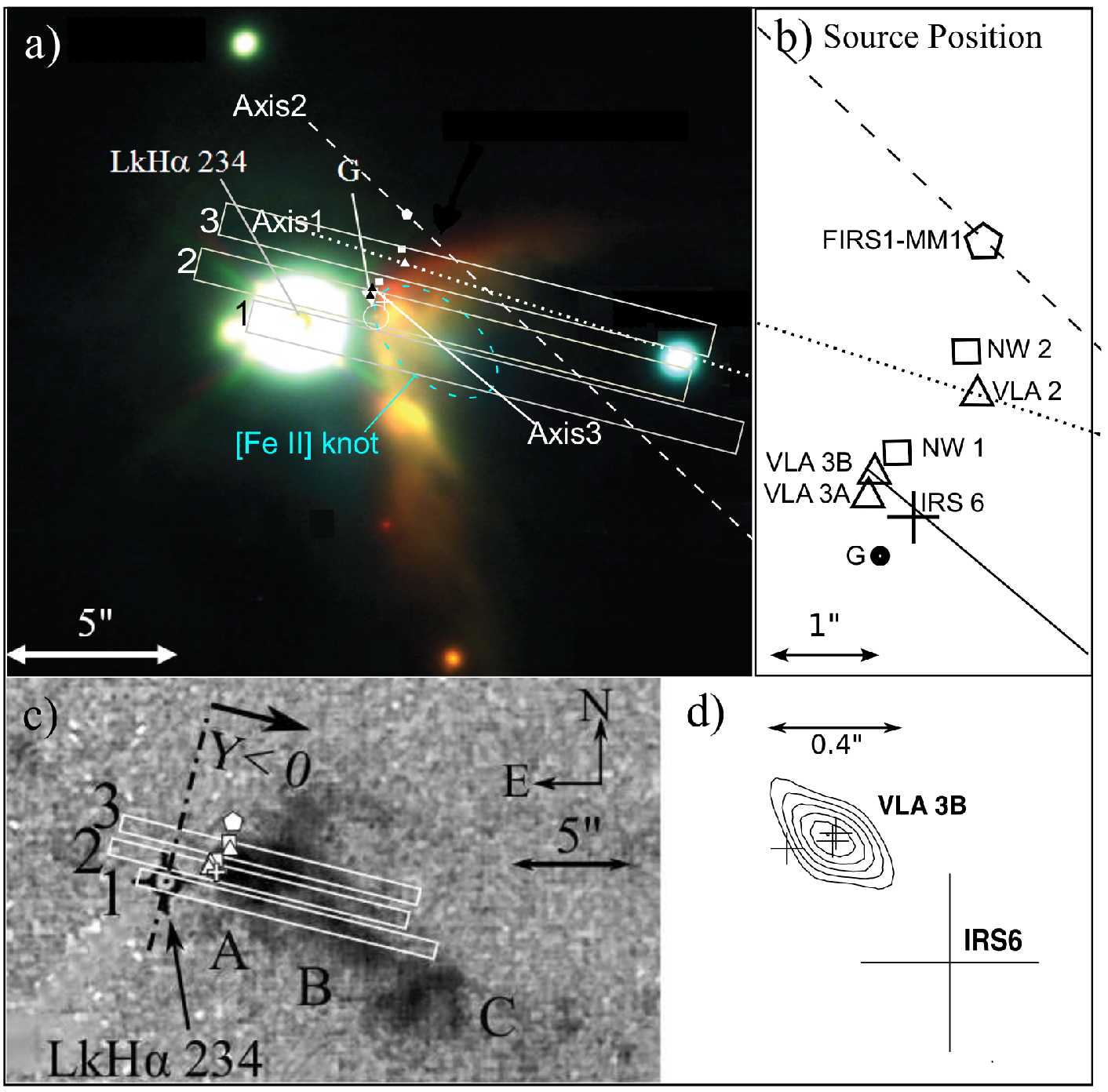}
\caption{Key for the Lk{\Ha} 234 region. a) Three slit positions, outflow source candidates, and axes of outflows. b) Zoom-in of source positions. The 1\farcs0 (W) $\times$ 14\farcs8 (L) slits at position angles (P.A.) of 256$\degree$ are superposed on the JHK color composite image \citep{Kato2011}. The slit positions are numbered as 1, 2 and 3 (SP 1, 2, and 3) from the bottom to the top in the figure. c) The slit positions on the continuum subtracted H$_{2}$ emission line image \citep{Cabrit1997}. A dash-dotted line in the inset represents the position of Y = 0$\arcsec$, and knots A, B, and C are shown. d) Source subtracted 3.6 cm map taken from \citet{Trinidad2004}.
FIRS1-MM1, VLA sources, mid-IR sources NW 1, NW 2 and IRS 6 are marked and a dashed ellipse shows the position of the bright {\FeII} knot detected at SP 1 and 2. The positional uncertainties of the sources are described in Section \ref{sec:intro}. The dotted, dashed, and solid lines marked as Axis1, Axis2, and Axis3 indicate axes of an optical {\SII} jet, a near-IR H$_{2}$ jet, and an {\FeII} jet which is newly suggested in this study, respectively. Axis1 and Axis2 correspond to A1 and A2 of \citet{Fuente2001}. Letter G is marked in the original image of \citet{Kato2011}, indicating the YSO candidate. \label{fig:sourcemap}}
\end{figure}

\begin{figure}
\epsscale{.7}
\plotone{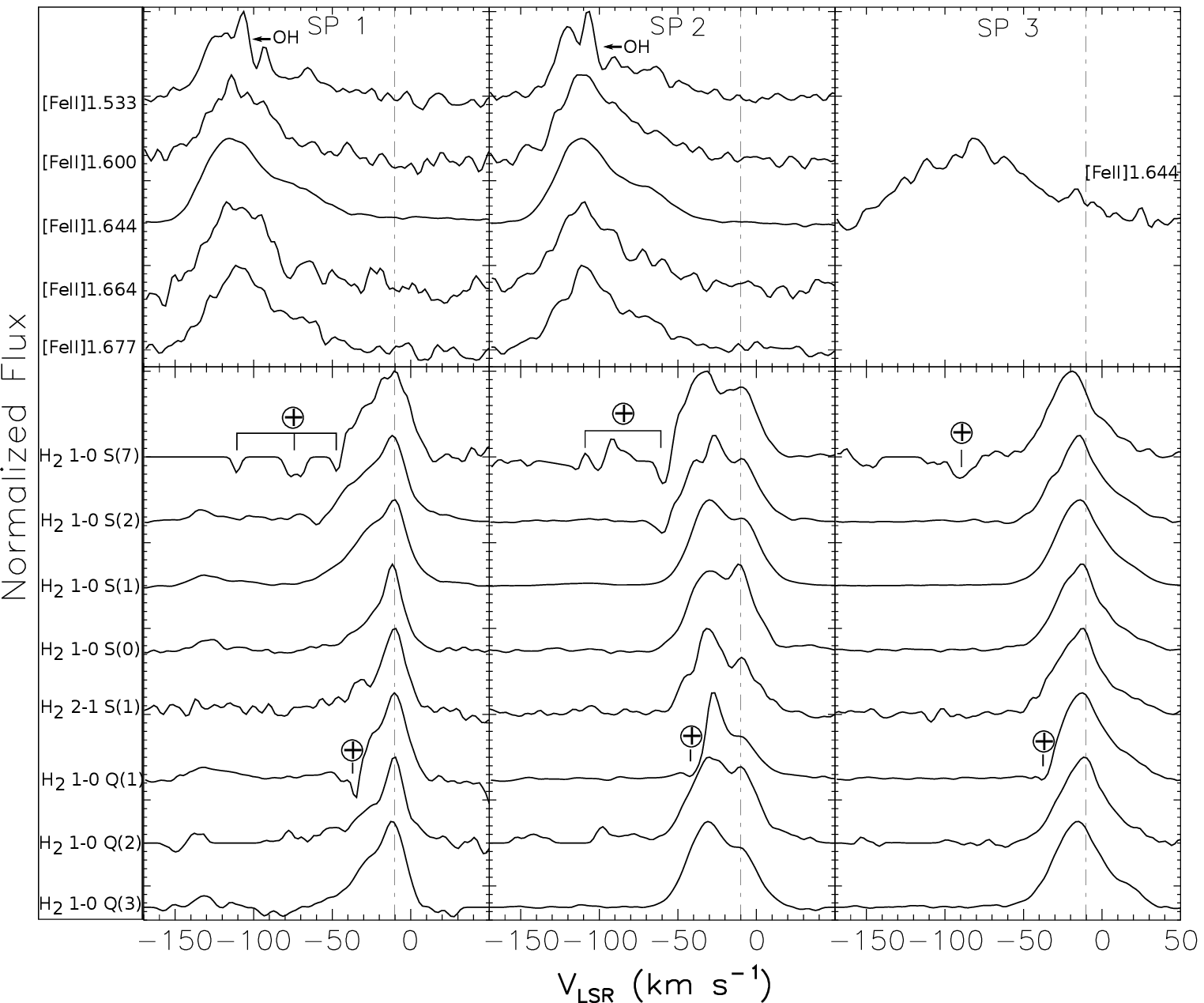}
\caption{Line profiles of {\FeII} and H$_{2}$ toward knot A at SP 1, 2, and 3. The fluxes are integrated over $-$6\farcs5 $<$ Y $<$ $-$0\farcs5. Upper three panels show the {\FeII} $\lambda$1.533 $\micron$, $\lambda$1.600 $\micron$, $\lambda$1.644 $\micron$, $\lambda$1.664 $\micron$, and $\lambda$1.677 $\micron$ lines. The bottom three panels show H$_{2}$ 1$-$0 S(7) $\lambda$1.748 $\micron$, 1$-$0 S(2) $\lambda$2.034 $\micron$, 1$-$0 S(1) $\lambda$2.122 $\micron$, 1$-$0 S(0) $\lambda$2.223 $\micron$, 2$-$1 S(1) $\lambda$2.247 $\micron$, 1$-$0 Q(1) $\lambda$2.407 $\micron$, 1$-$0 Q(2) $\lambda$2.413 $\micron$ and 1$-$0 Q(3) $\lambda$2.424 $\micron$ lines.
In each line, the flux is normalized to its peak value, and the profile is smoothed by gaussian filtering with $\sigma$ = 1 pixel. In the {\FeII} $\lambda$1.533 $\micron$ line profiles, we marked the locations where the line profiles were affected by the residual after OH sky subtraction. The `$\earth$' symbols indicate the telluric absorption features.
The systemic velocity at V$_{LSR}$ = $-$10.2 {\kms} \citep{Liu2011} is indicated with dash-dotted lines. \label{fig:profiles}}
\end{figure}

As shown in Section \ref{sec:intro} and Figure \ref{fig:sourcemap}, the Lk{\Ha} 234 region is complicated with several outflow sources observed at mid-IR, millimeter, and radio wavelengths. In this section, we summarize three separate outflows and their source candidates, and relate them to our own spectroscopic observations.

\begin{figure}
\epsscale{1.0}
\plotone{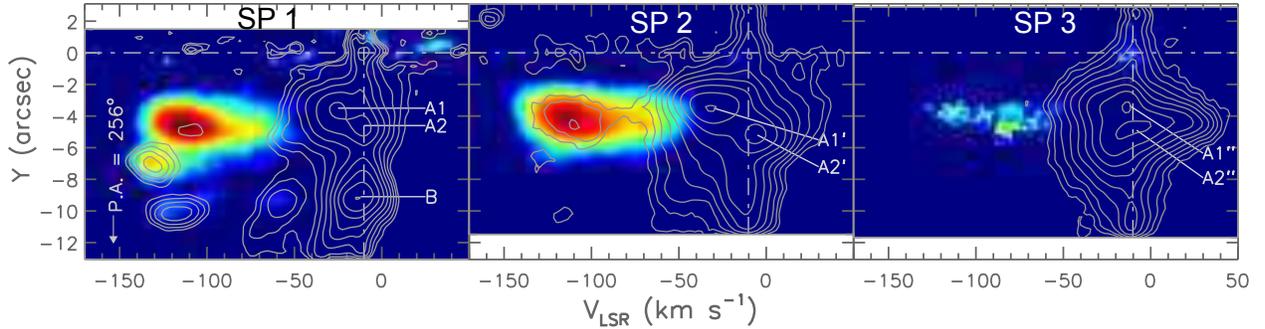}
\caption{PVDs of H$_{2}$ 1$-$0 S(1) $\lambda$2.122 $\micron$ (white contours) and {\FeII} $\lambda$1.644 $\mu$m (color intensity map) emission. The left, center, right panels correspond to slit position (SP) 1, 2, and 3, respectively. The contour starts from a 5$\sigma$ (1$\sigma$ = 0.02 $\times$ 10$^{-18}$ W m$^{-2}$ \AA$^{-1}$) level, and it increases with equal intervals in a logarithmic scale. In each panel, the highest contour level correspond to 2.6, 7.7, and 3.1 $\times$ 10$^{-18}$ W m$^{-2}$ \AA$^{-1}$. Continuum emission of Lk{\Ha} 234 is removed from SP 1 and 2. The horizontal and vertical dash-dotted  lines indicate Y = 0$\arcsec$ and systemic velocity (V$_{LSR}$ = $-$10.2 {\kms}), respectively. SP 1 and 2 cover knot A and B, and SP 3 only include knot A  (see Figure \ref{fig:sourcemap}c). Knot A shows multiple peak positions, thus we marked it as `A1' and `A2' in SP 1, `A1$\arcmin$' and `A2$\arcmin$' in SP 2, `A1$\arcsec$' and `A2$\arcsec$' in SP 3. Knot B is marked as `B' in SP 1. Northeast is up and southwest is down in the diagrams. \label{fig3}}
\end{figure}

\begin{figure}
\epsscale{1.0}
\plotone{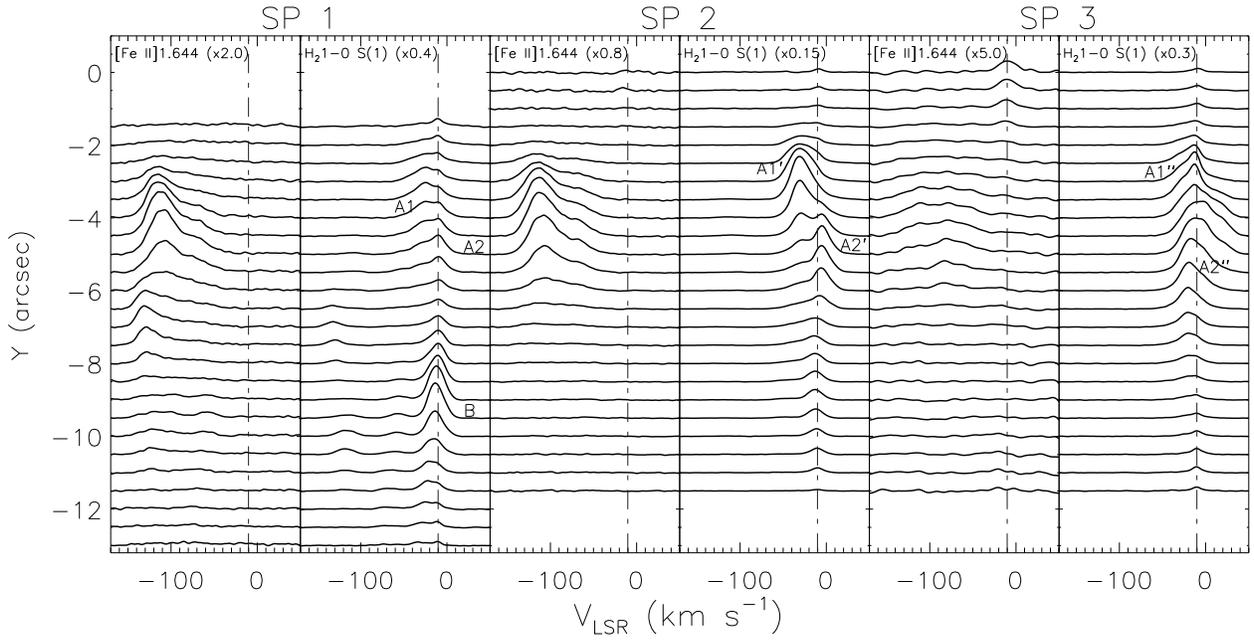}
\caption{Line profiles of {\FeII} $\lambda$1.644 $\micron$ and H$_{2}$ 1$-$0 S(1) $\lambda$2.122 $\micron$ lines in SP 1, 2, and 3. Line profiles are shown in every 0\farcs5 interval in Y-direction. The peak names indicated in Figure. \ref{fig3} are marked. The intensity levels are amplified with the factors of 2.0, 0.8, and 5.0 for {\FeII} emission, and 0.4, 0.15, and 0.3 for H$_{2}$ emission, respectively. The systemic velocity of V$_{LSR}$ = $-$10.2 {\kms} is indicated with dash-dotted lines. \label{fig4}}
\end{figure}

\begin{figure}
\epsscale{1.0}
\plotone{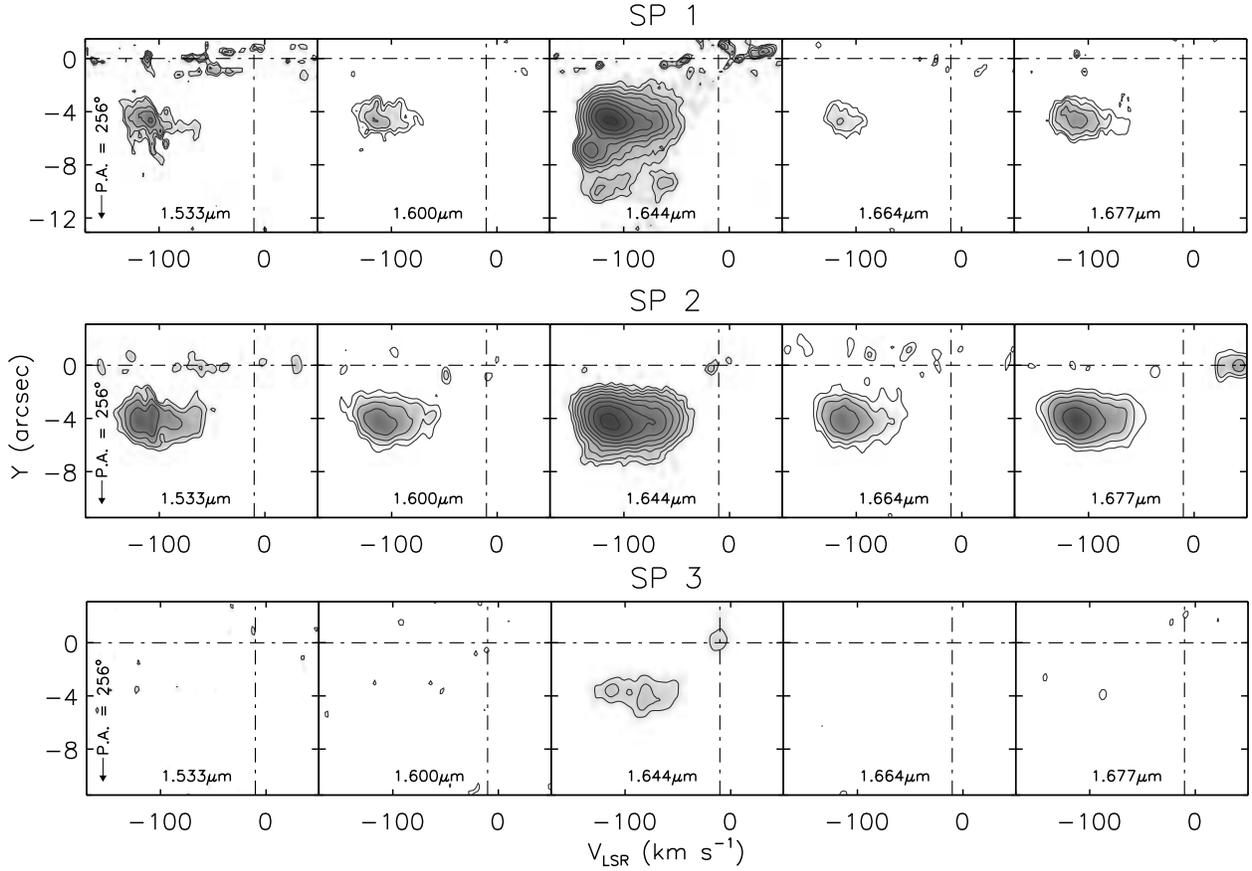}
\caption{PVDs of {\FeII} lines obtained from (top) SP 1, (middle) SP 2, and (bottom) SP 3. Contour starts from 4$\sigma$ level and it increases in equal intervals in a logarithmic scale. The highest contours in 1.644 $\mu$m indicate 0.6, 1.3, and 0.1 $\times$ 10$^{-18}$ W m$^{-2}$ \AA$^{-1}$ in SP 1, SP 2, and SP 3 respectively. In 1.533 and 1.600 $\mu$m, sky OH emission caused the absorption/emission features at $\sim$ $-$120 \kms. A dash-dotted horizontal line indicates Y = 0$\arcsec$ position. The systemic velocity is indicated with a dash-dotted vertical line in each panel. \label{fig:feiiPVD}}
\end{figure}

\begin{figure}
\epsscale{0.5}
\plotone{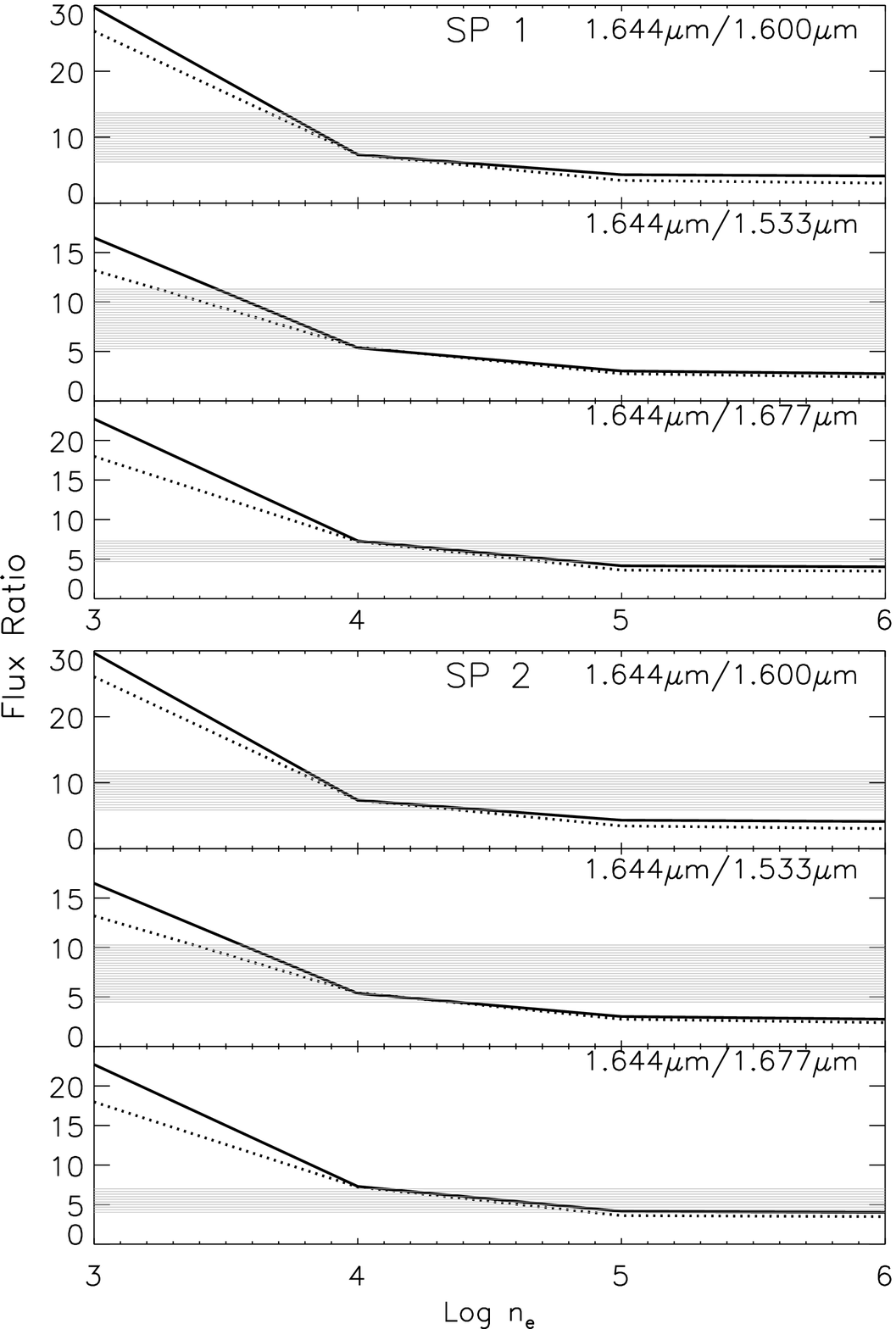}
\caption{The observed flux ratios of {\FeII} $\lambda$1.644 $\micron$/$\lambda$1.600 $\micron$, $\lambda$1.644 $\micron$/$\lambda$1.533 $\micron$, and $\lambda$1.644 $\micron$/$\lambda$1.677 $\micron$ in (top three panels) SP 1 and (bottom three panels) SP 2 superposed on the electron density ($n_{e}$) model from \citet{Nisini2002}. Solid and dotted lines represent excitation temperature of 4,000 K and 15,000 K in the model, respectively. The areas in grey -shade correspond to observed flux ratios including uncertainties. \label{fig:Ne}}
\end{figure}

\begin{figure}
\epsscale{.7}
\plotone{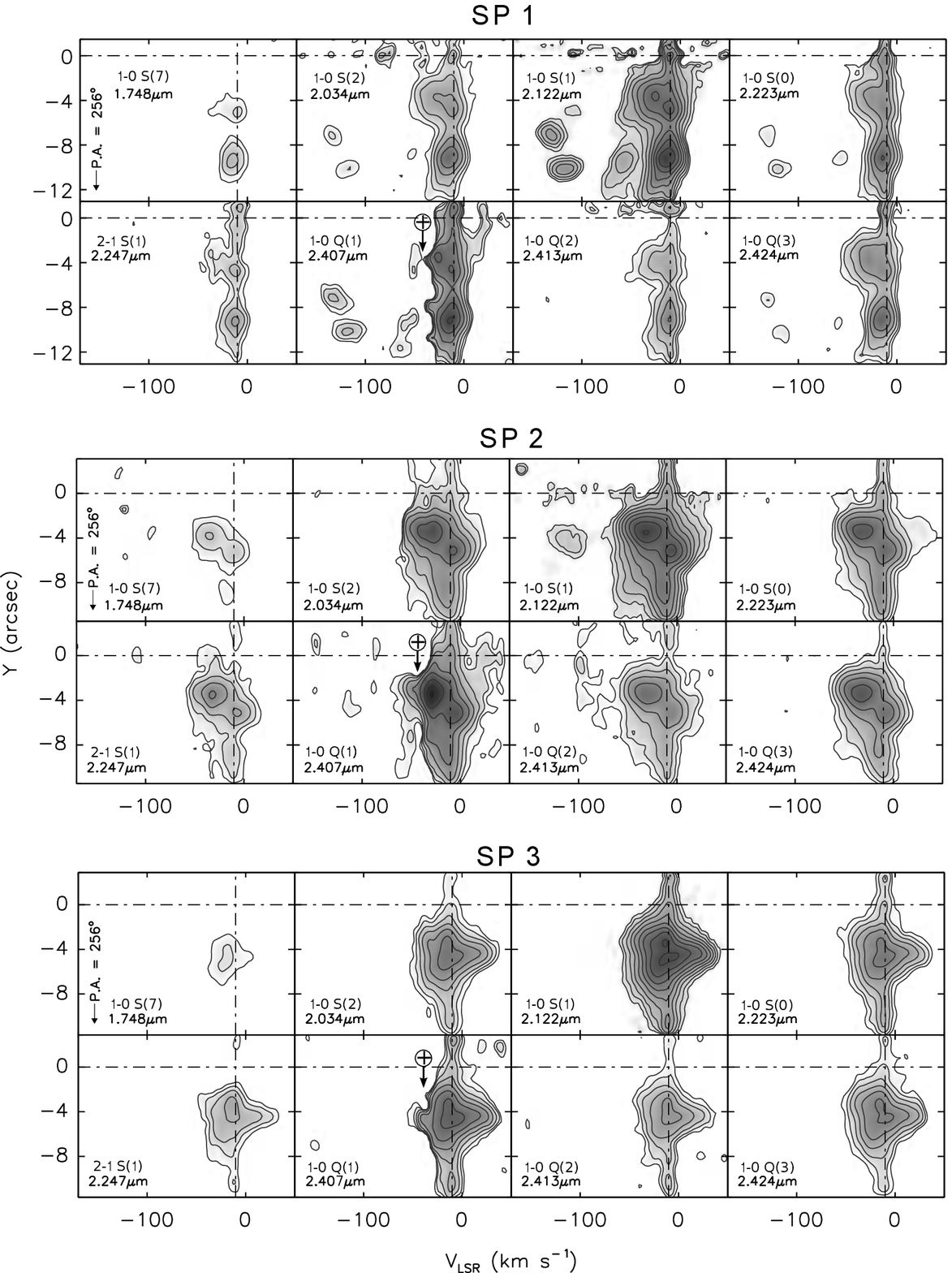}
\caption{PVDs of H$_{2}$ lines from (top) SP 1, (middle) SP 2, and (bottom) SP 3. The continuum emission from Lk{\Ha} 234 is subtracted in SP 1 and SP 2. The residuals from the subtraction are shown at Y = 0 $\pm$ 2$\arcsec$. Contour starts from 4$\sigma$ level and it increases in equal intervals in a logarithmic scale. The Highest contour levels in 2.122 $\mu$m correspond to 2.6, 7.7, and 3.1 $\times$ 10$^{-18}$ W m$^{-2}$ \AA$^{-1}$ for SP 1, 2, and 3. The `$\earth$' symbols at $-$40 {\kms} in 2.407 $\mu$m of all slit positions indicate the deep atmospheric absorption. Dash-dotted horizontal lines indicate the Y = 0$\arcsec$ position. The systemic velocity is shown with a dash-dotted vertical line in each panel. \label{fig:H2PVD}}
\end{figure}

\begin{figure}
\epsscale{.4}
\plotone{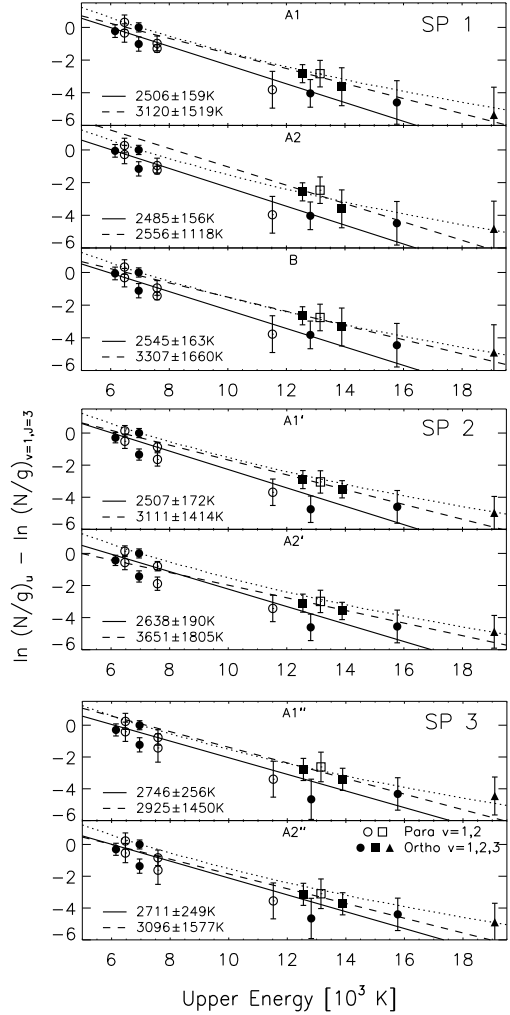}
\caption{The level population diagrams of H$_{2}$ for (top) knot A1, 2, and B of SP 1, (middle) A1$\arcmin$ and A2$\arcmin$ of SP 2, and (bottom) A1$\arcsec$ and A2$\arcsec$ of SP 3. Open and filled symbols indicate ortho and para transitions, respectively. Circles, squares, and triangles correspond to the upper vibrational levels of 1, 2, and 3 respectively. Solid and dashed lines are showing fitting lines for the vibrational levels of 1 and 2 from a weighted least-squares regression, respectively. Dotted lines represent the population ratio of the bow shock model \citep{Smith1991}. The excitation temperatures measured from the slopes of the lines are indicated in the lower-left of each panel. The A$_{\rm V}$ = 3.4 is used for reddening correction, and the values are normalized to 1$-$0 S(1) transition. \label{fig:H2pop}}
\end{figure}

1. The {\FeII} jet from VLA 3 (NW 1, IRS 6): 
At SP 1 and 2, we detected the bright {\FeII} emission with a velocity of $\sim$ $-$120 {\kms} at the position of knot A (Figure \ref{fig3}). Here, we suggest that this {\FeII} emission possibly originates from the outflow driven by radio source VLA 3B, according to following evidence.

The PVDs at SP 1 and 2 show that the {\FeII} knot lies between Y $=$ $-$2$\arcsec$ and Y $=$ $-$6$\arcsec$. As shown in Figure \ref{fig:sourcemap}, its position agrees well with the ``green blob'' feature seen in the center of the C-shaped reflection nebula in the JHK color image of \citet{Kato2011}, a feature that is probably caused by {\FeII} emission in the H$-$band. The narrow band {\FeII} $\lambda$1.644 $\micron$ image in \citet{Schultz1995}, which showed emission at the same position, supports this conclusion. The configuration of an {\FeII} jet detected along the central axis of a cavity in a reflection nebula is very similar to the case of L1551-IRS 5 \citep{Hayashi2009}. \citet{Kato2011} showed two jet-like features in their J$-$ and H$-$band images and they mentioned that those features point at YSO candidate G. They suggested that the jet-like feature containing the green blob is associated with the optical {\SII} jet.

We suggest another possible origin of the {\FeII} emission in Figure \ref{fig3}, VLA 3B.
The 3.6 cm continuum map of \citet{Trinidad2004} showed that VLA 3B has an elongated thermal radio jet extending northeast-southwest. In Figure \ref{fig:sourcemap}d, their source-subtracted 3.6 cm map is shown. We estimated a P.A. of $\sim$ 230$\degree$ for this jet. ``Axis3'' in Figure \ref{fig:sourcemap} denotes the axis of the radio jet in VLA 3B. This line passes through the positions of the {\FeII} knot detected in our observations and the green blob in the JHK image. The very similar major axis position angles of the {\FeII} emission in \citet{Schultz1995} and the radio thermal jet also support our suggestion, although the reliability of their {\FeII} image close to Lk{\Ha} 234 should be a cause of caution. Taking all factors into account, however, we conclude that this {\FeII} emission is from a jet driven by VLA 3B, with a velocity of V$_{LSR}$ $\sim$ $-$120 {\kms}. \citet{Trinidad2004} suggested that VLA 3B could be a low-mass YSO, while \citet{Kato2011} suggested NW 1 as a B6 $-$ B7 type YSO from the SED. An outflow cavity is opening toward the southwest from VLA 3B. The mid-IR source IRS 6 lies in the cavity with a $\sim$ 0\farcs5 offset from VLA 3B. The offset implies that the mid-IR emission comes from a region less affected by dust extinction.

 
We note that the P.A. of the {\FeII} jet is very similar to that of the H$_{2}$ jet (227$\degree$), as shown in Figure \ref{fig:sourcemap}. This may be an additional example of multiple jets sharing a similar orientation \citep[e.g.,][]{Nisini2001}, as pointed out by \citet{Trinidad2004}.

2. The H$_{2}$ jet from FIRS1-MM1: 
The most likely source of the H$_{2}$ jet \citep{Cabrit1997} and inner {\SII} jet \citep{Ray1990} is FIRS1-MM1 \citep{Fuente2001}, which is indicated as a white pentagon in Figure \ref{fig:sourcemap}. More accurately, it seems driving the knot B and C, because the position of FIRS1-MM1 is well-aligned with the line through those two knots (P.A. $\sim$ 227$\degree$), which is marked as ``Axis2'' in the Figure \ref{fig:sourcemap}. In Figure \ref{fig3} at SP 1, the H$_{2}$ emission shows a peak at Y = $-$9\farcs2. This peak position represents that SP 1 is superposed on the center position of knot B of H$_{2}$ jet, as we confirm in Figure \ref{fig:sourcemap}. We used our high-resolution observation to reveal, for the first time, that knot B shows multiple velocity peaks, which imply the presence of a bow shock (see also Section \ref{sec:4.1}).

3. The outer {\SII} jet from VLA 2 (NW 2): The proper motion and the locations of H$_{2}$O masers around radio continuum VLA 2 \citep{Trinidad2004,Marvel2005,Torrelles2014} confirmed that VLA 2 is the most likely source of redshifted CO outflow \citep{Mitchell1994,Fuente2001} and the outer {\SII} jet \citep[knot D $-$ E in][]{Ray1990}, showing good agreement in direction. In the Figure, the axis of {\SII} outflow is shown as ``Axis1'' with P.A. of $\sim$ 252$\degree$. Although SP 3 is located along the direction of the outer {\SII} jet from VLA 2 at Y $\sim$ $-$2\farcs6, we could not obtain information related to the outer {\SII} jet because there is no detection of any significant {\FeII} emission feature along SP 3, and the slit position only covers the inner {\SII} jet region \citep{Ray1990}.

\section{Summary} \label{sec:summary}

We have presented results from the first high-resolution near-IR spectroscopy toward the multiple outflows around the Herbig Be star Lk{\Ha} 234.

1. We detected 14 H$_{2}$ and 5 {\FeII} emission lines in the H$-$ and K$-$bands, including the H$_{2}$ 1$-$0 S(1) $\lambda$2.122 $\micron$ and $\lambda$1.644 $\micron$ lines.

2. We newly revealed an {\FeII} jet driven by radio source VLA 3B.

3. The multiple velocity peaks we observe in H$_{2}$ emission lines are consistent with a generic bow shock model. Both knots A and B show this bow shock feature. Furthermore, the positional difference ($\sim$ 1$\arcsec$) between low- and high-velocity components may be caused by a difference between the wing and apex of the bow.

4. The molecular hydrogen emission is dominant at low-velocity with a radial velocity within 50 {\kms} of the systemic velocity, while {\FeII} emission is only presnet in the higher-velocity ranges (V$_{LSR}$ $\sim$ $-$100 {\kms} to $-$130 {\kms}). This trend has been observed in previous studies and may be understood from the assumption that the H$_{2}$ emission originates from shocked gas entrained by a fast outflow. We also detected narrow H$_{2}$ emission along the whole slit length at systemic velocity, which indicate the background PDR.

5. From the ratios of {\FeII} emission lines, we estimate an electron density of $\sim$ 1.1 $\times$ 10$^{4}$ cm$^{-3}$. This is similar to or slightly smaller than the values in outflows from T-Tauri stars or Class 0$-$I sources.

6. The population diagrams of hydrogen molecules show that the gas is thermalized at a temperature of 2,500 $-$ 3000 K in this region. This indicates that the emission arises from shock excited gas. The ratios between H$_{2}$ lines at knot A and B are close to those from C-type shock models \citep{Smith1995}.

\acknowledgments
This work used the Immersion Grating Infrared Spectrograph (IGRINS) that was developed under a collaboration between the University of Texas at Austin and the Korea Astronomy and Space Science Institute (KASI) with the financial support of the US National Science Foundation under grant AST-1229522, of the University of Texas at Austin, and of the Korean GMT Project of KASI.


\clearpage


\clearpage
\begin{table}
\begin{center}
\caption{{\FeII} Line Fluxes Normalized to the 1.644 $\mu$m Flux.\label{tbl:feiiflux}}
\begin{tabular}{cccc}
\tableline\tableline
Line & $\lambda$($\mu$m)&\multicolumn{1}{c}{SP 1}&\multicolumn{1}{c}{SP 2}\\
\tableline
$a^{4}D_{5/2} - a^{4}F_{9/2}$ &1.53389 &0.12 $\pm$ 0.05 &0.14 $\pm$ 0.06\\
$a^{4}D_{3/2} - a^{4}F_{7/2}$ &1.59991 &0.10 $\pm$ 0.04 &0.11 $\pm$ 0.04\\
$a^{4}D_{7/2} - a^{4}F_{9/2}$ &1.64400 &1.00 $\pm$ 0.29 &1.00 $\pm$ 0.18\\
$a^{4}D_{1/2} - a^{4}F_{5/2}$ &1.66422 &0.09 $\pm$ 0.03 &0.10 $\pm$ 0.04\\
$a^{4}D_{5/2} - a^{4}F_{7/2}$ &1.67733 &0.17 $\pm$ 0.04 &0.18 $\pm$ 0.06\\
\tableline
\end{tabular}
\tablecomments{The reddening is not corrected.}
\end{center}
\end{table}

\clearpage

\begin{deluxetable}{ccccccccc}
\tabletypesize{\scriptsize}
\tablewidth{0pt}
\tablecaption{H$_{2}$ Line Fluxes Normalized to the 1$-$0 S(1) Flux.\label{tbl:H2flux}}
\tablehead{&&\multicolumn{3}{c}{SP 1}&\multicolumn{2}{c}{SP 2}&\multicolumn{2}{c}{SP 3}\\
\cmidrule(rl){3-5} \cmidrule(rl){6-7}\cmidrule(rl){8-9}
Line & $\lambda$($\mu$m) & A1&A2&B&A1\arcmin&A2\arcmin&A1\arcsec&A2\arcsec}
\startdata
1$-$0 S(9) & 1.68772 & 0.02 $\pm$ 0.01 & 0.02 $\pm$ 0.01 & 0.02 $\pm$ 0.01 & 0.02 $\pm$ 0.01 & 0.02 $\pm$ 0.01 & 0.02 $\pm$ 0.01 &  0.02 $\pm$ 0.01 \\
1$-$0 S(7) & 1.74803 & 0.04 $\pm$ 0.02 & 0.04 $\pm$ 0.02 & 0.05 $\pm$ 0.01 & 0.02 $\pm$ 0.01 & 0.02 $\pm$ 0.01 & 0.02 $\pm$ 0.01 &  0.02 $\pm$ 0.02 \\
1$-$0 S(6) & 1.78795 & 0.02 $\pm$ 0.01 & 0.02 $\pm$ 0.01 & 0.02 $\pm$ 0.01 & 0.02 $\pm$ 0.01 & 0.03 $\pm$ 0.01 & 0.03 $\pm$ 0.02 &  0.03 $\pm$ 0.02 \\
1$-$0 S(2) & 2.03376 & 0.19 $\pm$ 0.04 & 0.19 $\pm$ 0.04 & 0.19 $\pm$ 0.02 & 0.21 $\pm$ 0.03 & 0.23 $\pm$ 0.04 & 0.23 $\pm$ 0.06 &  0.22 $\pm$ 0.05 \\
2$-$1 S(3) & 2.07351 & 0.07 $\pm$ 0.04 & 0.07 $\pm$ 0.04 & 0.09 $\pm$ 0.01 & 0.08 $\pm$ 0.02 & 0.07 $\pm$ 0.03 & 0.09 $\pm$ 0.03 &  0.06 $\pm$ 0.03 \\
1$-$0 S(1) & 2.12183 & 1.00 $\pm$ 0.14 & 1.00 $\pm$ 0.13 & 1.00 $\pm$ 0.05 & 1.00 $\pm$ 0.14 & 1.00 $\pm$ 0.14 & 1.00 $\pm$ 0.14 &  1.00 $\pm$ 0.15 \\
2$-$1 S(2) & 2.15423 & 0.04 $\pm$ 0.02 & 0.06 $\pm$ 0.02 & 0.04 $\pm$ 0.01 & 0.03 $\pm$ 0.01 & 0.03 $\pm$ 0.01 & 0.05 $\pm$ 0.02 &  0.03 $\pm$ 0.02 \\
3$-$2 S(3) & 2.20139 & 0.01 $\pm$ 0.01 & 0.02 $\pm$ 0.01 & 0.02 $\pm$ 0.01 & 0.02 $\pm$ 0.01 & 0.02 $\pm$ 0.01 & 0.03 $\pm$ 0.02 &  0.02 $\pm$ 0.02 \\
1$-$0 S(0) & 2.22329 & 0.23 $\pm$ 0.05 & 0.22 $\pm$ 0.05 & 0.24 $\pm$ 0.02 & 0.19 $\pm$ 0.03 & 0.20 $\pm$ 0.04 & 0.22 $\pm$ 0.05 &  0.21 $\pm$ 0.05 \\
2$-$1 S(1) & 2.24771 & 0.08 $\pm$ 0.02 & 0.11 $\pm$ 0.03 & 0.10 $\pm$ 0.01 & 0.08 $\pm$ 0.02 & 0.06 $\pm$ 0.02 & 0.09 $\pm$ 0.03 &  0.06 $\pm$ 0.03 \\
1$-$0 Q(1) & 2.40659 & 0.40 $\pm$ 0.08 & 0.47 $\pm$ 0.08 & 0.47 $\pm$ 0.03 & 0.37 $\pm$ 0.06 & 0.32 $\pm$ 0.06 & 0.37 $\pm$ 0.07 &  0.37 $\pm$ 0.07 \\
1$-$0 Q(2) & 2.41344 & 0.14 $\pm$ 0.04 & 0.15 $\pm$ 0.04 & 0.14 $\pm$ 0.02 & 0.12 $\pm$ 0.03 & 0.11 $\pm$ 0.03 & 0.13 $\pm$ 0.04 &  0.11 $\pm$ 0.04 \\
1$-$0 Q(3) & 2.42373 & 0.27 $\pm$ 0.06 & 0.24 $\pm$ 0.05 & 0.25 $\pm$ 0.02 & 0.20 $\pm$ 0.03 & 0.18 $\pm$ 0.04 & 0.22 $\pm$ 0.05 &  0.19 $\pm$ 0.05 \\
1$-$0 Q(4) & 2.43749 & 0.08 $\pm$ 0.01 & 0.09 $\pm$ 0.01 & 0.07 $\pm$ 0.01 & 0.06 $\pm$ 0.01 & 0.05 $\pm$ 0.01 & 0.07 $\pm$ 0.03 &  0.06 $\pm$ 0.03 \\

\enddata
\tablecomments{The reddening is not corrected.}
\end{deluxetable}

\clearpage
\begin{deluxetable}{ccccccccccc}
\tabletypesize{\tiny}
\tablewidth{0pt}
\tablecaption{H$_{2}$ Line Ratios.\label{tbl:H2ratio}}
\tablehead{&\multicolumn{3}{c}{SP 1}&\multicolumn{2}{c}{SP 2}&\multicolumn{2}{c}{SP 3}&\multicolumn{3}{c}{Model}\\
\cmidrule(rl){2-4} \cmidrule(rl){5-6}\cmidrule(rl){7-8}\cmidrule(rl){9-11}
 & &&&&&&&C-&J-&UV\\
Ratio & A1&A2&B&A1\arcmin&A2\arcmin&A1\arcsec&A2\arcsec &shock&shock&Pumped}
\startdata
2$-$1 S(1) / 1$-$0 S(1) & 0.08 $\pm$ 0.02 & 0.11 $\pm$ 0.03 & 0.10 $\pm$ 0.01  & 0.08  $\pm$ 0.02 & 0.06 $\pm$ 0.02 & 0.09 $\pm$ 0.03 & 0.06 $\pm$ 0.03 & 0.05 & 0.24 & 0.55 \\
1$-$0 S(2) / 1$-$0 S(0) & 0.81 $\pm$ 0.23 & 0.86 $\pm$ 0.24 & 0.82 $\pm$ 0.10 & 1.07 $\pm$ 0.18 & 1.16 $\pm$ 0.26 & 1.07  $\pm$ 0.34 & 1.02 $\pm$ 0.34 & 1.56 & 2.08 & 1.09 \\
2$-$1 S(1) / 2$-$1 S(3) & 1.18 $\pm$ 0.74 & 1.50 $\pm$ 0.93 & 1.05 $\pm$ 0.17 & 0.97 $\pm$ 0.35 & 0.87$\pm$ 0.43 & 0.99  $\pm$ 0.47 & 0.97  $\pm$ 0.62 & 1.08 & 0.73 & 1.58 \\
1$-$0 S(1) / 1$-$0 Q(1) & 2.51 $\pm$ 0.49 & 2.11 $\pm$ 0.36 & 2.12 $\pm$ 0.15 & 2.67 $\pm$ 0.55 & 3.09 $\pm$ 0.43 & 2.69 $\pm$ 0.51 & 2.71 $\pm$ 0.52 & 1.29 & 1.59 & 1.00
\enddata
\tablecomments{The H$_{2}$ line ratios of C-, J- type shocks, and UV Pumped cases are taken from Table 6 of \citet{Beck2008}, who referred the calculations in \citet{Smith1995} and \citet{Black1987}.}

\end{deluxetable}




\end{document}